\documentclass[apj]{emulateapj}

\def\lapp{\ifmmode\stackrel{<}{_{\sim}}\else$\stackrel{<}{_{\sim}}$\fi}
\def\gapp{\ifmmode\stackrel{>}{_{\sim}}\else$\stackrel{>}{_{\sim}}$\fi}

\usepackage{natbib}
\usepackage{graphicx}
\usepackage{float}
\usepackage{color}

\submitted{Submitted to ApJL September 30, 2013; Accepted October 22, 2013}

\shorttitle{Ultra-stripped Type Ic SNe from close binaries}
\shortauthors{Tauris~et.~al.}

\begin{document}

\title{ULTRA-STRIPPED TYPE~Ic SUPERNOVAE FROM CLOSE BINARY EVOLUTION}

\author{
T.~M.~Tauris\altaffilmark{1,2}, N.~Langer\altaffilmark{1}, T.~J.~Moriya\altaffilmark{3,4}, Ph.~Podsiadlowski\altaffilmark{5}, S.-C.~Yoon\altaffilmark{6}, S.~I.~Blinnikov\altaffilmark{7,8,9}
}

\altaffiltext{1}{Argelander-Institut f\"ur Astronomie, Universit\"at Bonn, Auf dem H\"ugel 71, 53121 Bonn, Germany}
\altaffiltext{2}{Max-Planck-Institut f\"ur Radioastronomie, Auf dem H\"ugel 69, 53121 Bonn, Germany}
\altaffiltext{3}{Kavli Institute for the Physics and Mathematics of the Universe (WPI),
Todai Institutes for Advanced Study,
University of Tokyo, Kashiwanoha 5-1-5, Kashiwa, Chiba 277-8583, Japan}
\altaffiltext{4}{Research Center for the Early Universe, Graduate School of Science, University of Tokyo,
Hongo 7-3-1, Bunkyo, Tokyo 113-0033, Japan}
\altaffiltext{5}{Department of Astronomy, Oxford University, Oxford OX1~3RH, UK}
\altaffiltext{6}{Department of Physics \& Astronomy, Seoul National University, Gwanak-gu, Gwanak-ro 1, Seoul 151-742, Korea}
\altaffiltext{7}{Institute for Theoretical and Experimental Physics, Bolshaya Cheremushkinskaya 25, 117218 Moscow, Russia}
\altaffiltext{8}{Novosibirsk State University, Novosibirsk 630090, Russia}
\altaffiltext{9}{Sternberg Astronomical Institute, M.V.Lomonosov Moscow State University, Universitetski pr. 13, 119992 Moscow, Russia}

\begin{abstract}
Recent discoveries of weak and fast optical transients raise the question of their origin. 
We investigate the minimum ejecta mass associated with core-collapse supernovae (SNe) of Type~Ic.
We show that mass transfer from a helium star to a compact companion can produce an ultra-stripped core which 
undergoes iron core collapse and leads to an extremely fast and faint SN~Ic. In this Letter, 
a detailed example is presented in which the pre-SN stellar mass is barely 
above the Chandrasekhar limit, resulting in the ejection of only $\sim\!0.05-0.20\;M_{\odot}$ 
of material and the formation of a low-mass neutron star (NS). We compute synthetic light curves
of this case and demonstrate that SN~2005ek could be explained by our model. 
We estimate that the fraction of such ultra-stripped to all SNe could be as high as $10^{-3}-10^{-2}$.
Finally, we argue that the second explosion in some double NS systems  
(for example, the double pulsar PSR~J0737$-$3039B) was likely associated with an ultra-stripped SN~Ic.
\end{abstract}

\email{tauris@astro.uni-bonn.de}

\section{Introduction}\label{sec:intro}
In recent years, high-cadence surveys and dedicated supernova (SN) searches have increased the discovery rate of 
unusual optical transients. Examples of events with low peak 
luminosities ($\la 10^{42}\;{\rm erg}\,{\rm s}^{-1}$) and rapidly decaying light curves 
comprise SN~2005ek \citep{dsm+13}, SN~2010X \citep{kkg+10} and SN~2005E \citep{pgm+10}.
These SNe show diverse spectroscopic signatures \citep[cf.][]{kk13} and thus 
their explanation may require a diversity of models. \citet{dsm+13} concluded that
SN~2005ek represents the most extreme small ratio of ejecta to remnant mass observed for a
core-collapse SN to date.

Two classes of models have been suggested to explain these fast and faint events.
The first one proposes a partial explosion \citep{bswn07} or the collapse of a white dwarf \citep{dbo+06},
both involving a very low ejecta mass. The second relates
to hydrogen-free core-collapse SNe with large amounts of fall back \citep{mtt+10} 
or with essentially no radioactive nickel \citep{dhl+11,kk13}.

Also the afterglows of short $\gamma$--ray bursts \citep[sGRBs,][]{ffp+05,ber10},  
which are suggested to originate from mergers of neutron stars 
\citep[NSs; e.g.][]{bnpp84,pac86,elps89}, are predicted theoretically  
to produce fast and faint ``kilonovae'' at optical/IR wavelengths powered by 
the ejection of $\sim\!0.01\;M_{\odot}$ of radioactive $r$-process material
\citep{lp98b,mmd+10,bk13} -- see \citet{bfc13,tlf+13} for detections
of a possible candidate. These mergers constitute the prime candidate sources of high frequency gravitational waves 
to be detected by the LIGO/VIRGO network within the next 5~yr \citep{aaa+13},
which fosters the search for their electromagnetic counterparts \citep[e.g.][]{kn13}.
A good understanding of faint and fast transients is thus urgently needed.

It is well-known that the outcome of stellar evolution in close binaries differs significantly from that of single stars.
The main effects of mass loss/gain, and tidal forces at work, are changes in the stellar rotation rate, the nuclear burning scheme and
the wind mass-loss rate \citep{lan12}.
As a result, the binary interactions affect the final core mass prior to collapse \citep{bhl+01,plp+04},
and therefore the type of compact remnant left behind and the amount of envelope mass ejected.

Whereas most Type~Ib/c SNe are expected to originate in binary systems, from the initially more massive star
which has been stripped of its hydrogen envelope by mass-transfer to its companion \citep{eit08}, 
these pre-SN stars typically have an envelope mass of $1\;M_{\odot}$ or more \citep{ywl10}. 
In a close X-ray binary, however, a second mass-transfer stage from a helium star to a NS 
can strip the helium star further prior to the SN \citep[][and references therein]{dpsv02,dp03,ibk+03}. 
The fast decay of the Type~Ic SN~1994I with an absolute magnitude of $M_{\rm V}\approx -18$ was explained by \citet{nyp+94} 
from such a model, and for which they suggested a resulting pre-SN carbon-oxygen star of
$\sim\!2\;M_{\odot}$ and a corresponding ejecta mass of $\sim\!0.9\;M_{\odot}$.
The starting point of all these calculations (tight systems containing a naked helium star and a NS with orbital period, $P_{\rm orb}< 2\;{\rm days}$)
is a continuation of the expected outcome of common envelope evolution (CE) in high-mass X-ray binaries.
Of particular interest is mass transfer by Roche-lobe overflow (RLO) initiated by the 
helium star expansion after core helium exhaustion (so-called Case~BB RLO).

Here, we investigate SN progenitors originating from the evolution of such helium star--NS binaries, and find
them to be the most promising candidates to achieve maximally stripped pre-SN cores.
We provide a detailed example and demonstrate that this scenario can produce an iron core collapse of
a small, bare core of $\sim\!1.5\;M_{\odot}$, leading to an ejecta mass of only $\sim\!0.1\;M_{\odot}$.
In Section~\ref{sec:binary_evol} we present our model with computations extending beyond oxygen ignition, and
study the final core structure prior to iron core collapse. We model light curves for the resulting fast and faint SN 
explosion and compare our results with SN~2005ek in Section~\ref{sec:obs}.
Finally, we estimate the rate of these events, discuss double NS binaries and summarize our conclusions in Section~\ref{sec:summary}.

\section{Binary evolution and formation of a nearly naked pre-SN metal core}\label{sec:binary_evol}
\begin{figure}[t]
\centering
\includegraphics[width=0.80\columnwidth,angle=-90]{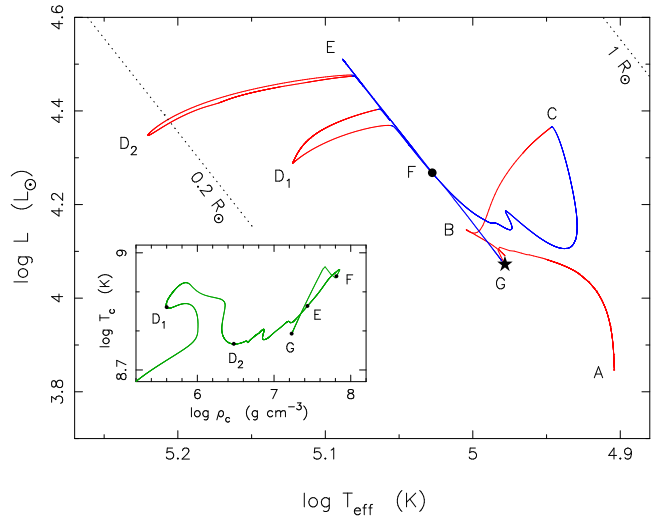}
\caption{
  H-R~diagram of the evolution of a $2.9\;M_{\odot}$ helium star which loses mass to a NS companion star prior to the core-collapse SN.
  The red part of the track corresponds to a detached system and the blue part is marking RLO.
  The evolutionary sequence A, B, C, D$_1$, D$_2$, E, F, G is explained in the text.
  The inner panel shows the final evolution in the central temperature--central density plane.
  }
\label{fig:HR}
\end{figure}

\begin{figure}[t]
\centering
\includegraphics[width=0.80\columnwidth,angle=-90]{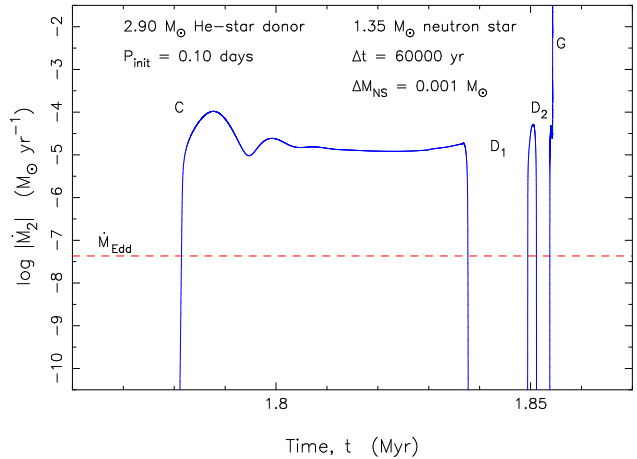}
\caption{
  Mass-transfer rate as a function of stellar age for the helium star evolution plotted in Figures~\ref{fig:HR}~and~\ref{fig:kippenhahn}. 
  The major phase of mass transfer is the Case~BB RLO lasting for about $56\,000\;{\rm yr}$.
  The mass-transfer rate is seen to be highly super-Eddington in this phase
  (the horizontal dashed line marks the Eddington accretion rate, $\dot{M}_{\rm Edd}$).
  The Case~BB RLO is followed by a detached phase for about $12\,000\;{\rm yr}$ because of core carbon burning (D$_1$).
  Following a second detached phase (D$_2$), a rigorous helium shell flash leads to the spike in point G -- see text. 
  }
\label{fig:Mdot}
\end{figure}

For our detailed calculations of the evolution of helium star--NS binaries,
we applied the BEC stellar evolution code \citep[][and references therein]{ywl10}; for recent applications 
to X-ray binaries and further description of Case~BB RLO, see e.g. \citet{tlk11,tlk12,ltk+13}.
We assumed an initial helium star donor mass of $M_{\rm He}=2.9\;M_{\odot}$,  
a NS mass of $M_{\rm NS}=1.35\;M_{\odot}$ and orbital period, $P_{\rm orb}=0.10^{\rm d}$.
Figure~\ref{fig:HR} shows the complete evolution of the helium star donor in the H-R~diagram. 
For the same model, we present the calculated mass-transfer rate, the Kippenhahn diagram and the final chemical structure
in Figures~\ref{fig:Mdot}--\ref{fig:abundances}. 

The points marked by letters along the evolutionary track in Figure~\ref{fig:HR} correspond to: 
A) helium star zero-age main sequence ($t=0$); 
B) core helium exhaustion at $t=1.75\;{\rm Myr}$, defining the bottom of the giant branch with shell helium burning;
C) onset of Case~BB RLO at $t=1.78\;{\rm Myr}$; 
D$_1$) core carbon burning during $t=1.836-1.849\;{\rm Myr}$, leading to radial contraction and Roche-lobe detachment
       (as a result of the mirror principle when the core expands);
D$_2$) consecutive ignitions of carbon burning shells during $t=1.850-1.854\;{\rm Myr}$ and Roche-lobe detachment again; 
E) maximum luminosity  at $t=1.854\,304\;{\rm Myr}$; 
F) off-center ($m/M_{\odot}\simeq 0.5$) ignition of oxygen burning at $t=1.854\,337\;{\rm Myr}$ 
($T_{\rm c}=9.1\times 10^8\;{\rm K}$; $\rho _{\rm c}=7.1\times 10^7\;{\rm g}\,{\rm cm}^{-3}$), marked by a bullet.
Shortly thereafter (point G) the binary orbit becomes dynamically unstable when $\dot{M}_2 >10^{-2}\;M_{\odot}\,{\rm yr}^{-1}$ 
and we took our model star out of the binary to continue its final evolution as an isolated star ($t\ge 1.854\,337\;{\rm Myr}$).
At this stage $P_{\rm orb}=0.070^{\rm d}$. 
Off-center oxygen burning ignites at $t=1.854\,553\;{\rm Myr}$ and 3--4~yr thereafter our computer code breaks down (see below).  

The evolution of our model is similar to those of \citet{dpsv02} (cf. their fig.~6), who calculated their
binary stellar models until carbon burning. Our model resembles also very well the cores of $9.5-11\;M_{\odot}$ single stars
which were found to lead to iron core collapse \citep{uyt12, jhn+13}.
From fig.~4e of the latter work, we infer that our core is expected to undergo iron core 
collapse $\sim\!10\;{\rm yr}$ after our calculations were terminated.

Figure~\ref{fig:Mdot} shows the mass-transfer rate, $|\dot{M}_2|$ as a function of time.
The total duration of the mass-transfer phases is seen to last for about $\Delta t =60\,000\;{\rm yr}$ (excluding
a couple of detached epochs), which causes the NS to accrete an amount
$\Delta M_{\rm NS}=(0.7-2.1)\times 10^{-3}\;M_{\odot}$, depending on the assumed accretion efficiency and the exact value of
the Eddington accretion limit, $\dot{M}_{\rm Edd}$. Here we assumed $\dot{M}_{\rm Edd}=3.9\times 10^{-8}\;M_{\odot}\,{\rm yr}^{-1}$
(a typical value for accretion of helium rich matter) and allowed for the actual accretion rate to be somewhere in the interval
30\%--100\% of this value \citep[see recent discussion by][]{ltk+13}.
We note that the structure of the donor star is hardly affected by uncertainties in the NS accretion efficiency.

The tiny helium burning layer causes the star to expand vigorously, 
approximately at the same time the convective shell penetrates to the surface (cf. point~G),
resulting in numerical problems for our code. We therefore end our calculations without resolving this flash which is, in any case, 
not important for the core structure
and the NS accretion due to the little amount of mass in the inflated envelope.
As discussed in detail by \citet{dp03,ibk+03}, the runaway mass transfer may lead to the onset of a CE evolution, 
where the final outcome is determined by the  
competition between the timescale for the onset of spiral-in and the remaining lifetime before gravitational collapse.
In general, the duration of the CE and spiral-in phase is found to be $<10^3\;{\rm yr}$ \citep[e.g.][]{pod01},
which is long compared to the estimated final life time of our model of about 10~yr \citep{jhn+13}.
Therefore, while our model may still lose even more helium, the naked star is expected to undergo iron core collapse
and produce a SN~Ic (see Section~\ref{subsec:LC} for a discussion of the SN type).

\subsection{Final Core Structure}\label{subsec:core}
\begin{figure}
\centering
\includegraphics[width=1.10\columnwidth]{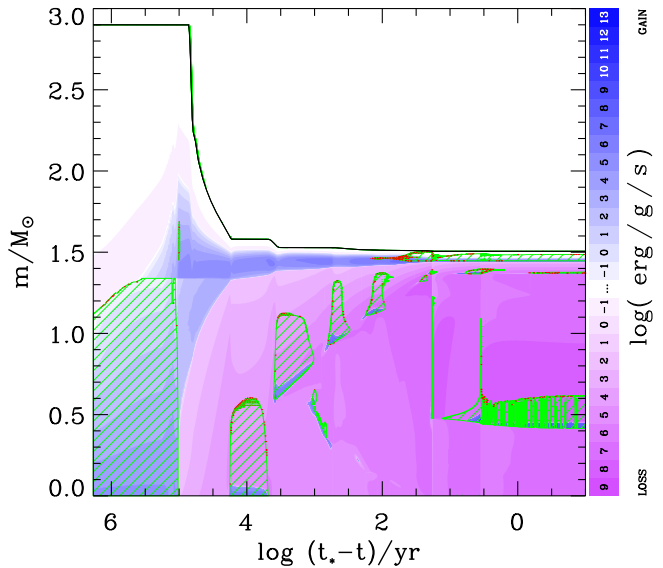}
\caption{
Kippenhahn diagram of the $2.9\,M_{\odot}$ helium star undergoing Case~BB/BBB RLO in 
Figures~\ref{fig:HR}--\ref{fig:Mdot}.
The plot shows cross-sections of the helium star in mass-coordinates
from the center to the surface of the star, along the $y$-axis, as a function of stellar age on the $x$-axis.
The value $(t_{*}-t)/{\rm yr}$ is the remaining time of our calculations, spanning a total time of $t_*= 1.854356~{\rm Myr}$.
The green hatched areas denote zones with convection; red color indicates semi-convection.
The intensity of the blue/purple color indicates the net energy-production rate.
Shortly after off-centered oxygen ignition (at $m/M_{\odot}\simeq 0.5$, when $\log (t_*-t)=1.3$)
we evolved the star further as an isolated star for $\sim\!20\,{\rm yr}$ until our code crashed,  
about 10~yr prior to core collapse.
}
\label{fig:kippenhahn}
\end{figure}

\begin{figure}
\centering
\includegraphics[width=0.80\columnwidth,angle=-90]{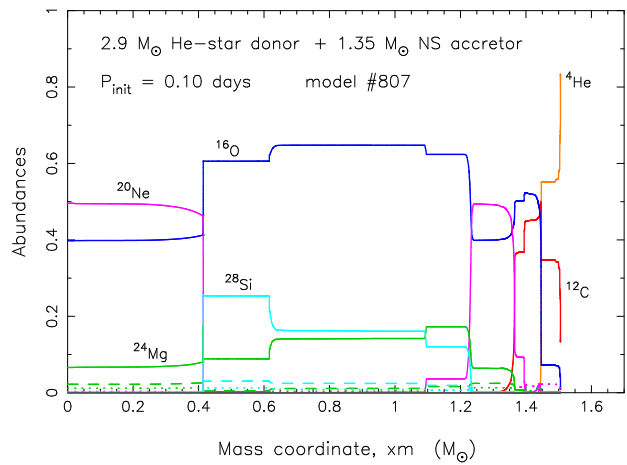}
\caption{
  chemical abundance structure of an ultra-stripped SN~Ic progenitor star (from our last calculated model \#807 at $t=1.854356\;{\rm Myr}$),
  This naked $1.50\;M_{\odot}$ pre-collapsing star has a hybrid structure with an ONeMg inner core enclothed by a thick OMgSi outer core, which again
  is enclothed by shells of ONeMg and CO, and outermost a tiny envelope with $0.033\;M_{\odot}$ of helium.
  }
\label{fig:abundances}
\end{figure}

In Figures~\ref{fig:kippenhahn}--\ref{fig:abundances} we present the final chemical structure of our stellar model,
whose final mass is $1.50\;M_{\odot}$. 
It consists of an almost naked metal core of $1.45\;M_{\odot}$ which is covered by a helium-rich envelope of only $0.05\;M_{\odot}$.
The Kippenhahn diagram in Figure~\ref{fig:kippenhahn} shows the interior structure and evolution (energy production and convection zones) of the 
$2.9\;M_{\odot}$ helium star which undergoes Case~BB RLO, and later Case~BBB RLO
(RLO re-initiated following shell carbon burning), and leaves behind a silicon-rich core. 
Seven carbon burning shells follow the core carbon burning phase. 
Oxygen ignites off-center (near a mass coordinate of $m/M_{\odot} \simeq 0.5$) 
due to an inversion of the temperature profile produced by neutrino cooling in the inner core.
The convection zone on top of the initial oxygen burning shell (at $\log (t_{*}-t) \simeq 1.3$)
reaches out to $m/M_{\odot} \simeq 1.2$. Hence, the final chemical structure of our model (Figure~\ref{fig:abundances}) consists 
of a thick OMgSi outer core of $\sim\!0.8\;M_{\odot}$, which is sandwiched by ONeMg-layers in the inner core ($\sim\!0.4\;M_{\odot}$) and 
toward the envelope ($0.14\;M_{\odot}$). This is surrounded by a CO-layer of $0.1\;M_{\odot}$ and a helium-rich envelope 
of $\sim\!0.05\;M_{\odot}$. The total amount of helium in this envelope is $0.033\;M_{\odot}$. 

The ultra-stripped nature of our model is achieved because the helium star is forced to lose (almost) its entire
envelope in a very tight orbit where a NS can fit in.

\section{Observational consequences}\label{sec:obs}
\subsection{SN Light Curves and Spectral Type}\label{subsec:LC}
\begin{figure}
\centering
\includegraphics[width=0.68\columnwidth,angle=-90]{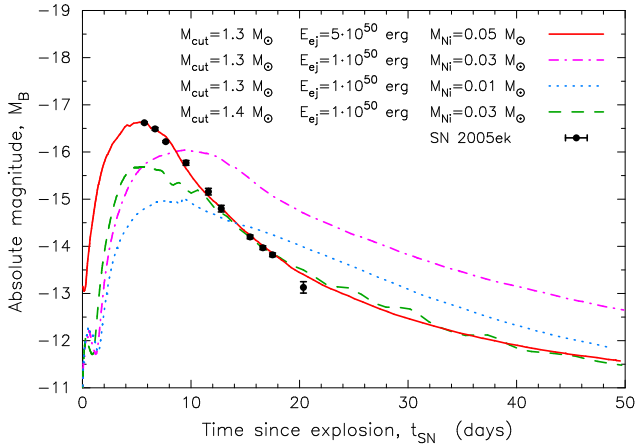}
\caption{
$B$-band SN light curves of ultra-stripped SNe~Ic from the progenitor calculation obtained by binary evolution.
The different curves correspond to various combinations of mass cut ($M_{\rm cut}$),
explosion energy ($E_{\rm ej}$) and amount of $^{56}$Ni synthesized ($M_{\rm Ni}$). The total bolometric
luminosities, $L_{\rm bol}$ of the four curves peak in the range $0.4-2.3\times 10^{42}\;{\rm erg}\,{\rm s}^{-1}$.
The data for SN~2005ek is taken from \citet{dsm+13}. The explosion date is arbitrarily chosen to match the light curve.
}
\label{fig:LC}
\end{figure}
\label{subsec:spectra}

We modeled the SN light curve evolution based on the SN progenitor star presented in Section~\ref{sec:binary_evol},
with SN kinetic energy, nickel mass and mass cut as free parameters. 
The light curve calculations were performed by a one-dimensional multi-group radiation hydrodynamics code \verb|STELLA| \citep[e.g.][]{brs+06}.
Figure~\ref{fig:LC} shows the obtained $B$-band light curves.
We applied two mass cuts ($M_{\rm cut}=1.3\;M_{\odot}$ and $1.4\;M_{\odot}$) to the progenitor star
with a corresponding SN ejecta mass ($M_{\rm ej}=M_{*}-M_{\rm cut}$) of $0.2\;M_{\odot}$ and $0.1\;M_{\odot}$, respectively.

As seen in Figure~\ref{fig:LC}, the calculation assuming a mass cut of $M_{\rm cut}=1.3\;M_{\odot}$, a nickel mass of
$M_{\rm Ni}=0.05\;M_{\odot}$ and a SN explosion energy of $E_{\rm ej}=5\times 10^{50}\;{\rm erg}$ agrees well with the 
extremely rapidly declining Type~Ic SN~2005ek \citep{dsm+13}.
The multi-color light curve evolution of our model also matches that of SN~2005ek.
These parameters are fairly consistent with those estimated by \citet{dsm+13} 
($M_{\rm ej}\sim\!0.3\;M_{\odot}$, $E_{\rm ej}\sim\!2.5\times 10^{50}\;{\rm erg}$, and $M_{\rm Ni}\sim\!0.03\;M_{\odot}$).
As the rise time ($t_{\rm rise}\propto M_{\rm ej}^{3/4}\, E_{\rm ej}^{-1/4}$, \citealt{arn82}) of SN~2005ek is not well-constrained by the
observations, $M_{\rm ej}$ and $E_{\rm ej}$ remain uncertain and we expect some degeneracy in the light curve models. 
Therefore, the set of parameters shown in Figure~\ref{fig:LC} for SN~2005ek may not be unique.

Recent modeling of synthetic SN spectra by \citet{hmt+12} suggests that more than $0.06\;M_{\odot}$ of helium is needed
for helium lines to become visible in optical/IR spectra. 
Our final stellar model contains only $0.033\;M_{\odot}$ of helium. This amount could be further reduced 
by mass transfer and winds, as well as by explosive helium burning during the SN explosion.
Therefore, the SN corresponding to our model is expected to be observed
as a SN~Ic, rather than a SN~Ib.

\subsection{Comparison to Regular SNe~Ib/c}
\label{subsec:regular}
\citet{ywl10} showed that the stellar radii of regular SNe~Ib/c progenitors in
binary systems are generally larger for lower final masses. This is
because the helium envelope expands farther for a more compact carbon-oxygen core
during the evolutionary stages beyond core helium exhaustion. 
Relatively low-mass progenitors of SNe~Ib/c with
such extended envelopes will appear systematically more luminous 
than more massive SNe~Ib/c progenitors \citep{ygv+12}. 

However the ultra-stripped SN~Ic progenitors considered in the
present study have a very compact nature at the pre-SN stage despite their
low final masses, due to the close proximity of the compact companion
and the resultant tiny amount of helium left in the envelope.
The final radius of about $0.4\;R_{\odot}$, bolometric luminosity of about $12\,000\;L_{\odot}$ and a corresponding high
surface temperature of about $95\,000\;{\rm K}$,
as predicted by our model (Figure~\ref{fig:HR}), imply that they are visually faint ($M_{\rm V} \simeq +1.9$), 
and it will be extremely difficult to identify them in pre-SN optical/IR images.
Depending on the column density of the accretion disk corona and the circumbinary material from 
the ejected envelope, along the line-of-sight to Earth, these ultra-stripped systems may be bright X-ray sources prior to the SN.

If the mass transfer ends before the core collapse, our stripped progenitor is not expected to develop an optically thick wind. 
Therefore, the photosphere of the shock breakout will form near the hydrostatic stellar surface.  
The corresponding X-ray flash will have a harder energy spectrum, $kT \approx 10\;{\rm keV}$, and a
shorter duration, $\delta t_{\rm s} \approx 30\;{\rm s}$ 
(e.g. according to recent simulations and analytical estimates by \citealt{tbn13} and \citealt{skw13}), than in the
case of regular SN~Ib/c (such as SN~2008D) for which $kT \approx 0.1\;{\rm keV}$ and 
$\delta t_{\rm s} \approx 200\;{\rm s}$ \citep[e.g.][]{sbp+08,mlb+09}. 
This is because the progenitors of regular SNe Ib/c have either a
larger radius or an optically thick wind, making the shock breakouts occur at
relatively low temperatures.  

\subsection{Rates of Ultra-stripped SNe~Ic}\label{subsec:rates}
The expected rate of ultra-stripped SNe~Ic as discussed above can be estimated as follows. The evolution of
a binary system must avoid a merger during the CE~phase which produces the helium star--NS binary,
but the resulting orbit must be close enough such that the compact star is able to
peel-off the helium shell during the mass transfer initiated by the helium star expansion.

Immediate descendants of such systems are close double NS binaries which eventually merge
due to gravitational wave radiation. The merger rate of those 
in a Milky~Way-like galaxy is estimated to be between $\sim$ a~few~Myr$^{-1}$ and
up to $\sim\!100\;{\rm Myr}^{-1}$ \citep[e.g.][and references therein]{aaa+10}.
Compared to a total Galactic SN rate of $\sim\!1-2$~per~century, this would mean that
only one in $10^2-10^4$~SNe is of this kind. 
On the other hand, we have to include similar systems producing black hole--NS binaries
and also add the number of systems (a factor of a few) which were disrupted as a consequence
of the dynamical effects of the second SN explosion, and are therefore not contributing to the estimated merger rates quoted above.

In addition, there are several other channels including a helium star and a relatively compact companion star that could also produce ultra-stripped SNe~Ic:
(1) helium star--white dwarf binaries and (2) the progenitors of intermediate-mass/low-mass X-ray binaries (LMXBs) and binary millisecond pulsars, where
the companion is a relatively compact main-sequence star. 
The binary pulsar PSR~J1141$-$6545 provides an example for the first of these evolutionary channels  
\citep[Case~BB RLO would occur between stages~7 and 8 in fig.~1, in][]{ts00}. The existence of a non-recycled (and thereby short-lived) pulsar
in this system with $P_{\rm orb}=0.20^{\rm d}$ suggests that the rate for this channel could be at least comparable to that of the double NS merger channel.
The progenitors of LMXBs with $P_{\rm orb}<1\;{\rm day}$ are also potential systems producing, at least partly, stripped SNe~Ic. Their birthrate is probably
$\la 10^{-5}\;{\rm yr}^{-1}$ \citep{prp03}, but since the vast majority of these systems become disrupted in the SN (i.e. before the observed LMXB phase),
the rate for stripped SNe~Ic from this channel could be up to a factor of 10 higher. 
We note that the stripping effect is less certain for a giant helium star experiencing mass transfer to a low-mass main-sequence star and more research is needed to investigate this.

While it is difficult to estimate the rate of ultra-stripped SNe~Ic, these arguments taken together suggest that a realistic estimate of the 
ratio of ultra-stripped SNe~Ic to the total SN rate is likely to be in the range of $10^{-2}-10^{-3}$. 

\subsection{Resulting Neutron Star Birth Properties}\label{subsec:NSmass}
The extreme stripping producing the bare pre-SN core investigated here leads to an iron 
core-collapse SN with little ejecta mass, $M_{\rm ej}$. It is expected that
the NS left behind will have a gravitational mass in the range $1.18-1.31\;M_{\odot}$, 
depending on $M_{\rm cut}$ and the yet unknown equation-of-state (EoS) of NS matter and its associated release of gravitational binding energy \citep{ly89}. 
These values are in accordance with the NS masses measured in some double NS systems \citep[e.g.][]{ksm+06}.
Including the uncertainties in $M_{\rm cut}$ and the EoS, the total range of NS birth masses produced in iron core-collapse SNe may thus range from $1.10-1.70\;M_{\odot}$
\citep[see][for a discussion of the upper limit]{tlk11}.
The magnitude of any kick imparted to the newborn NS from an ultra-stripped iron core-collapse SN is uncertain. It may depend on, for example, $M_{\rm cut}$ (and $E_{\rm ej}$) and thus on the
timescale of the explosion compared to those of the non-radial hydrodynamic instabilities producing large kicks \citep[e.g.][]{plp+04,pdl+05,jan12}.
For a symmetric explosion, the post-SN eccentricity of our system will be $e\simeq 0.07-0.13$, also in agreement with constraints obtained from some of the known double NS systems.
In case our modeled pre-SN core had a slightly smaller mass, the outcome would have been an electron capture SN \citep{plp+04,pdl+05}.
In general, we conclude that an ultra-stripped core, following an evolution similar to the one presented here, is an evident SN~Ic progenitor candidate of any second iron core-collapse 
or electron capture SN forming a close-orbit double NS system. The young radio pulsar PSR~J0737$-$3039B, in the double pulsar system \citep{ksm+06}, is an example of a NS 
which is most likely to have formed this way.

\section{Conclusions}\label{sec:summary}
We have shown that post-CE mass stripping in a
helium star--NS binary can produce a SN progenitor 
with a total mass of only $\sim\! 1.50\;M_{\odot}$ and an envelope mass of barely $0.05\;M_{\odot}$.
The resulting iron core collapse of our example sequence leads to a Type~Ic SN with an ejecta mass 
in the range $0.05-0.20\;M_{\odot}$. We estimate that one in every 100--1000 SNe may be of this ultra-stripped type.
Through synthetic light curve calculations, 
we have demonstrated that SN~2005ek is a viable candidate for such an event. 
Given the current ambitious observational efforts to search for peculiar and weak optical transients,
it seems probable that more such ultra-stripped SN~Ic with diminutive ejecta will be detected within the coming years. 
Finally, we conclude that ultra-stripped cores are evident SN~Ic progenitors of both 
iron core-collapse and electron capture SNe producing the second NS in close-orbit double NS systems.


\newpage
\begin{thebibliography}{50}

\bibitem[{{Aasi} {et~al.}(2013){Aasi}, {Abadie}, {Abbott}, {Abbott}, {Abbott},
  {Abernathy}, {Accadia}, {Acernese}, \& et~al.}]{aaa+13}
{Aasi}, J., {Abadie}, J., {Abbott}, B.~P., {et~al.} 2013, ArXiv
  astro-ph:1304.0670

\bibitem[{{Abadie} {et~al.}(2010){Abadie}, {Abbott}, {Abbott}, {Abernathy},
  {Accadia}, {Acernese}, {Adams}, {Adhikari}, {Ajith}, {Allen}, \&
  et~al.}]{aaa+10}
{Abadie}, J., {Abbott}, B.~P., {Abbott}, R., {et~al.} 2010, Classical and
  Quantum Gravity, 27, 173001

\bibitem[{{Arnett}(1982)}]{arn82}
{Arnett}, W.~D. 1982, \apj, 253, 785

\bibitem[{{Barnes} \& {Kasen}(2013)}]{bk13}
{Barnes}, J., \& {Kasen}, D. 2013, \apj, 775, 18

\bibitem[{{Berger}(2010)}]{ber10}
{Berger}, E. 2010, \apj, 722, 1946

\bibitem[{{Berger} {et~al.}(2013){Berger}, {Fong}, \& {Chornock}}]{bfc13}
{Berger}, E., {Fong}, W., \& {Chornock}, R. 2013, \apjl, 774, L23

\bibitem[{{Bildsten} {et~al.}(2007){Bildsten}, {Shen}, {Weinberg}, \&
  {Nelemans}}]{bswn07}
{Bildsten}, L., {Shen}, K.~J., {Weinberg}, N.~N., \& {Nelemans}, G. 2007,
  \apjl, 662, L95

\bibitem[{{Blinnikov} {et~al.}(1984){Blinnikov}, {Novikov}, {Perevodchikova},
  \& {Polnarev}}]{bnpp84}
{Blinnikov}, S.~I., {Novikov}, I.~D., {Perevodchikova}, T.~V., \& {Polnarev},
  A.~G. 1984, Soviet Astronomy Letters, 10, 177

\bibitem[{{Blinnikov} {et~al.}(2006){Blinnikov}, {R{\"o}pke}, {Sorokina},
  {Gieseler}, {Reinecke}, {Travaglio}, {Hillebrandt}, \&
  {Stritzinger}}]{brs+06}
{Blinnikov}, S.~I., {R{\"o}pke}, F.~K., {Sorokina}, E.~I., {et~al.} 2006, \aap,
  453, 229

\bibitem[{{Brown} {et~al.}(2001){Brown}, {Heger}, {Langer}, {Lee}, {Wellstein},
  \& {Bethe}}]{bhl+01}
{Brown}, G.~E., {Heger}, A., {Langer}, N., {et~al.} 2001, New Astronomy, 6,
  457

\bibitem[{{Dessart} {et~al.}(2006){Dessart}, {Burrows}, {Ott}, {Livne}, {Yoon},
  \& {Langer}}]{dbo+06}
{Dessart}, L., {Burrows}, A., {Ott}, C.~D., {et~al.} 2006, \apj, 644, 1063

\bibitem[{{Dessart} {et~al.}(2011){Dessart}, {Hillier}, {Livne}, {Yoon},
  {Woosley}, {Waldman}, \& {Langer}}]{dhl+11}
{Dessart}, L., {Hillier}, D.~J., {Livne}, E., {et~al.} 2011, \mnras, 414, 2985

\bibitem[{{Dewi} \& {Pols}(2003)}]{dp03}
{Dewi}, J.~D.~M., \& {Pols}, O.~R. 2003, \mnras, 344, 629

\bibitem[{{Dewi} {et~al.}(2002){Dewi}, {Pols}, {Savonije}, \& {van den
  Heuvel}}]{dpsv02}
{Dewi}, J.~D.~M., {Pols}, O.~R., {Savonije}, G.~J., \& {van den Heuvel},
  E.~P.~J. 2002, \mnras, 331, 1027

\bibitem[{{Drout} {et~al.}(2013){Drout}, {Soderberg}, {Mazzali}, {Parrent},
  {Margutti}, {Milisavljevic}, {Sanders}, {Chornock}, {Foley}, {Kirshner},
  {Filippenko}, {Li}, {Brown}, {Cenko}, {Chakraborti}, {Challis}, {Friedman},
  {Ganeshalingam}, {Hicken}, {Jensen}, {Modjaz}, {Perets}, {Silverman}, \&
  {Wong}}]{dsm+13}
{Drout}, M.~R., {Soderberg}, A.~M., {Mazzali}, P.~A., {et~al.} 2013, \apj, 774,
  58

\bibitem[{{Eichler} {et~al.}(1989){Eichler}, {Livio}, {Piran}, \&
  {Schramm}}]{elps89}
{Eichler}, D., {Livio}, M., {Piran}, T., \& {Schramm}, D.~N. 1989, \nat, 340,
  126

\bibitem[{{Eldridge} {et~al.}(2008){Eldridge}, {Izzard}, \& {Tout}}]{eit08}
{Eldridge}, J.~J., {Izzard}, R.~G., \& {Tout}, C.~A. 2008, \mnras, 384, 1109

\bibitem[{{Fox} {et~al.}(2005){Fox}, {Frail}, {Price}, {Kulkarni}, {Berger},
  {Piran}, {Soderberg}, {Cenko}, {Cameron}, {Gal-Yam}, {Kasliwal}, {Moon},
  {Harrison}, {Nakar}, {Schmidt}, {Penprase}, {Chevalier}, {Kumar}, {Roth},
  {Watson}, {Lee}, {Shectman}, {Phillips}, {Roth}, {McCarthy}, {Rauch},
  {Cowie}, {Peterson}, {Rich}, {Kawai}, {Aoki}, {Kosugi}, {Totani}, {Park},
  {MacFadyen}, \& {Hurley}}]{ffp+05}
{Fox}, D.~B., {Frail}, D.~A., {Price}, P.~A., {et~al.} 2005, \nat, 437, 845

\bibitem[{{Hachinger} {et~al.}(2012){Hachinger}, {Mazzali}, {Taubenberger},
  {Hillebrandt}, {Nomoto}, \& {Sauer}}]{hmt+12}
{Hachinger}, S., {Mazzali}, P.~A., {Taubenberger}, S., {et~al.} 2012, \mnras,
  422, 70

\bibitem[{{Ivanova} {et~al.}(2003){Ivanova}, {Belczynski}, {Kalogera}, {Rasio},
  \& {Taam}}]{ibk+03}
{Ivanova}, N., {Belczynski}, K., {Kalogera}, V., {Rasio}, F.~A., \& {Taam},
  R.~E. 2003, \apj, 592, 475

\bibitem[{{Janka}(2012)}]{jan12}
{Janka}, H.-T. 2012, Annual Review of Nuclear and Particle Science, 62, 407

\bibitem[{{Jones} {et~al.}(2013){Jones}, {Hirschi}, {Nomoto}, {Fischer},
  {Timmes}, {Herwig}, {Paxton}, {Toki}, {Suzuki}, {Mart{\'{\i}}nez-Pinedo},
  {Lam}, \& {Bertolli}}]{jhn+13}
{Jones}, S., {Hirschi}, R., {Nomoto}, K., {et~al.} 2013, \apj, 772, 150

\bibitem[{{Kasliwal} \& {Nissanke}(2013)}]{kn13}
{Kasliwal}, M.~M., \& {Nissanke}, S. 2013, ArXiv astro-ph:1309.1554

\bibitem[{{Kasliwal} {et~al.}(2010){Kasliwal}, {Kulkarni}, {Gal-Yam}, {Yaron},
  {Quimby}, {Ofek}, {Nugent}, {Poznanski}, {Jacobsen}, {Sternberg}, {Arcavi},
  {Howell}, {Sullivan}, {Rich}, {Burke}, {Brimacombe}, {Milisavljevic},
  {Fesen}, {Bildsten}, {Shen}, {Cenko}, {Bloom}, {Hsiao}, {Law}, {Gehrels},
  {Immler}, {Dekany}, {Rahmer}, {Hale}, {Smith}, {Zolkower}, {Velur},
  {Walters}, {Henning}, {Bui}, \& {McKenna}}]{kkg+10}
{Kasliwal}, M.~M., {Kulkarni}, S.~R., {Gal-Yam}, A., {et~al.} 2010, \apjl, 723,
  L98

\bibitem[{{Kleiser} \& {Kasen}(2013)}]{kk13}
{Kleiser}, I., \& {Kasen}, D. 2013, ArXiv astro-ph:1309.4088

\bibitem[{{Kramer} {et~al.}(2006){Kramer}, {Stairs}, {Manchester},
  {McLaughlin}, {Lyne}, {Ferdman}, {Burgay}, {Lorimer}, {Possenti}, {D'Amico},
  {Sarkissian}, {Hobbs}, {Reynolds}, {Freire}, \& {Camilo}}]{ksm+06}
{Kramer}, M., {Stairs}, I.~H., {Manchester}, R.~N., {et~al.} 2006, Science,
  314, 97

\bibitem[{{Langer}(2012)}]{lan12}
{Langer}, N. 2012, \araa, 50, 107

\bibitem[{{Lattimer} \& {Yahil}(1989)}]{ly89}
{Lattimer}, J.~M., \& {Yahil}, A. 1989, \apj, 340, 426

\bibitem[{{Lazarus} {et~al.}(2013){Lazarus}, {Tauris}, {Knispel}, \& {et
  al.}}]{ltk+13}
{Lazarus}, P., {Tauris}, T.~M., {Knispel}, B., \& {et al.} 2013, MNRAS, in press,
ArXiv astro-ph:1310.5857

\bibitem[{{Li} \& {Paczy{\'n}ski}(1998)}]{lp98b}
{Li}, L.-X., \& {Paczy{\'n}ski}, B. 1998, \apjl, 507, L59, L59

\bibitem[{{Metzger} {et~al.}(2010){Metzger}, {Mart{\'{\i}}nez-Pinedo},
  {Darbha}, {Quataert}, {Arcones}, {Kasen}, {Thomas}, {Nugent}, {Panov}, \&
  {Zinner}}]{mmd+10}
{Metzger}, B.~D., {Mart{\'{\i}}nez-Pinedo}, G., {Darbha}, S., {et~al.} 2010,
  \mnras, 406, 2650

\bibitem[{{Modjaz} {et~al.}(2009){Modjaz}, {Li}, {Butler}, \& {et
  al.}}]{mlb+09}
{Modjaz}, M., {Li}, W., {Butler}, N., \& {et al.} 2009, \apj, 702, 226

\bibitem[{{Moriya} {et~al.}(2010){Moriya}, {Tominaga}, {Tanaka}, {Nomoto},
  {Sauer}, {Mazzali}, {Maeda}, \& {Suzuki}}]{mtt+10}
{Moriya}, T., {Tominaga}, N., {Tanaka}, M., {et~al.} 2010, \apj, 719, 1445

\bibitem[{{Nomoto} {et~al.}(1994){Nomoto}, {Yamaoka}, {Pols}, {van den Heuvel},
  {Iwamoto}, {Kumagai}, \& {Shigeyama}}]{nyp+94}
{Nomoto}, K., {Yamaoka}, H., {Pols}, O.~R., {et~al.} 1994, \nat, 371, 227

\bibitem[{{Paczynski}(1986)}]{pac86}
{Paczynski}, B. 1986, \apjl, 308, L43

\bibitem[{{Perets} {et~al.}(2010){Perets}, {Gal-Yam}, {Mazzali}, {Arnett},
  {Kagan}, {Filippenko}, {Li}, {Arcavi}, {Cenko}, {Fox}, {Leonard}, {Moon},
  {Sand}, {Soderberg}, {Anderson}, {James}, {Foley}, {Ganeshalingam}, {Ofek},
  {Bildsten}, {Nelemans}, {Shen}, {Weinberg}, {Metzger}, {Piro}, {Quataert},
  {Kiewe}, \& {Poznanski}}]{pgm+10}
{Perets}, H.~B., {Gal-Yam}, A., {Mazzali}, P.~A., {et~al.} 2010, \nat, 465,
  322

\bibitem[{{Pfahl} {et~al.}(2003){Pfahl}, {Rappaport}, \&
  {Podsiadlowski}}]{prp03}
{Pfahl}, E., {Rappaport}, S., \& {Podsiadlowski}, P. 2003, \apj, 597, 1036

\bibitem[{{Podsiadlowski}(2001)}]{pod01}
{Podsiadlowski}, P. 2001, in Astronomical Society of the Pacific Conference
  Series, Vol. 229, Evolution of Binary and Multiple Star Systems, ed.
  {P.~Podsiadlowski, S.~Rappaport, A.~R.~King, F.~D'Antona, \& L.~Burderi },
  239

\bibitem[{{Podsiadlowski} {et~al.}(2005){Podsiadlowski}, {Dewi}, {Lesaffre},
  {Miller}, {Newton}, \& {Stone}}]{pdl+05}
{Podsiadlowski}, P., {Dewi}, J.~D.~M., {Lesaffre}, P., {et~al.} 2005, \mnras,
  361, 1243

\bibitem[{{Podsiadlowski} {et~al.}(2004){Podsiadlowski}, {Langer},
  {Poelarends}, {Rappaport}, {Heger}, \& {Pfahl}}]{plp+04}
{Podsiadlowski}, P., {Langer}, N., {Poelarends}, A.~J.~T., {et~al.} 2004, \apj,
  612, 1044

\bibitem[{{Sapir} {et~al.}(2013){Sapir}, {Katz}, \& {Waxman}}]{skw13}
{Sapir}, N., {Katz}, B., \& {Waxman}, E. 2013, \apj, 774, 79

\bibitem[{{Soderberg} {et~al.}(2008){Soderberg}, {Berger}, {Page}, \& {et
  al.}}]{sbp+08}
{Soderberg}, A.~M., {Berger}, E., {Page}, K.~L., \& {et al.} 2008, \nat, 453,
  469

\bibitem[{{Tanvir} {et~al.}(2013){Tanvir}, {Levan}, {Fruchter}, {Hjorth},
  {Hounsell}, {Wiersema}, \& {Tunnicliffe}}]{tlf+13}
{Tanvir}, N.~R., {Levan}, A.~J., {Fruchter}, A.~S., {et~al.} 2013, \nat, 500,
  547

\bibitem[{{Tauris} {et~al.}(2011){Tauris}, {Langer}, \& {Kramer}}]{tlk11}
{Tauris}, T.~M., {Langer}, N., \& {Kramer}, M. 2011, \mnras, 416, 2130

\bibitem[{{Tauris} {et~al.}(2012){Tauris}, {Langer}, \& {Kramer}}]{tlk12}
---. 2012, \mnras, 425, 1601

\bibitem[{{Tauris} \& {Sennels}(2000)}]{ts00}
{Tauris}, T.~M., \& {Sennels}, T. 2000, \aap, 355, 236

\bibitem[{{Tolstov} {et~al.}(2013){Tolstov}, {Blinnikov}, \&
  {Nadyozhin}}]{tbn13}
{Tolstov}, A.~G., {Blinnikov}, S.~I., \& {Nadyozhin}, D.~K. 2013, \mnras, 429,
  3181

\bibitem[{{Umeda} {et~al.}(2012){Umeda}, {Yoshida}, \& {Takahashi}}]{uyt12}
{Umeda}, H., {Yoshida}, T., \& {Takahashi}, K. 2012, 
Prog. Theor. Exp. Phys., DOI: 10.1093/ptep/pts017

\bibitem[{{Yoon} {et~al.}(2010){Yoon}, {Woosley}, \& {Langer}}]{ywl10}
{Yoon}, S., {Woosley}, S.~E., \& {Langer}, N. 2010, \apj, 725, 940

\bibitem[{{Yoon} {et~al.}(2012){Yoon}, {Gr{\"a}fener}, {Vink}, {Kozyreva}, \&
  {Izzard}}]{ygv+12}
{Yoon}, S.-C., {Gr{\"a}fener}, G., {Vink}, J.~S., {Kozyreva}, A., \& {Izzard},
  R.~G. 2012, \aap, 544, L11

\end{thebibliography}
\end{document}